\documentstyle[preprint,revtex]{aps}
\begin{document}
\draft
\preprint{HYUPT-93/08}
\begin{title}
Static Solution of the General Relativistic\\ 
Nonlinear $\sigma$-Model Equation
\end{title} 
\author{Chul H. Lee, Joon Ha Kim and Hyun Kyu Lee}
 
\begin{instit}
Department of Physics, Hanyang University, Seoul 133-791, Korea
\end{instit}
\begin{abstract}
The nonlinear $\sigma$-model is considered to be useful in describing 
hadrons (Skyrmions) in low energy hadron physics and the approximate
behavior of the global texture. Here we investigate the properties of
the static solution of the nonlinear $\sigma$-model equation coupled
with gravity. As in the case where gravity is ignored, there is still
no scale parameter that determines the size of the static solution and
the winding number of the solution is $1/2$. The geometry of the spatial
hyperspace in the asymptotic region of large $r$ is explicitly shown to 
be that of a flat space with some missing solid angle.
\end{abstract}
When a certain symmetry of a system is broken spontaneously, a masseless 
scalar particle naturally  emerges as a Goldstone boson\cite{goldstone}
for each broken generator of the symmetry. The simplest lagrangian density, 
which has 
been used to demonstrate the above formal argument, can be written as
 
\begin{eqnarray}
{\cal L}=\frac{1}{2}\partial^{\mu} \phi^{\alpha} \partial_{\mu} \phi^{\alpha} 
-\lambda(\mid\phi\mid^2 -v^2)^2
\label{lphi4}
\end{eqnarray}
where $\alpha = 0,1,2,3$ are chosen as an example in this paper. The physical
vacuum of the system can be defined as a field configuration  which satisfies
$\mid\phi\mid^2 =v^2$. If we choose the $\phi^0$-direction as the one that 
acquires the non-vanishing vacuum expectation value such that
 
\begin{eqnarray}
 \langle \phi^0 \rangle &=& v\nonumber\\
\langle \phi^i \rangle &=& 0 \  ,\ \ i= 1,2,3\label{vev}
\end{eqnarray}
then the symmetry of the system, $SO(4)$, is reduced to $SO(3)$ and 
$\phi^i$'s become 
masseless Goldstone bosons. In the limit, $\lambda \rightarrow \infty$, 
the massive mode in the $\phi^0$-direction becomes infinitely heavy and 
decoupled from the system. The symmetry now can be realized in terms of 
$\phi^i$'s only: 
the unbroken symmetry is realized linearly but the transformation in the 
direction of broken generators nonlinearly. 
 
In terms of Goldstone bosons, the simplest lagrangian density in which the 
whole symmetry of the system is properly
incorporated as discussed above is the nonlinear $\sigma$-model
lagrangian density
\begin{eqnarray} 
{\cal L}=-\frac{f_{\phi}^2}{4} {\rm Tr} (\partial^{\mu} U 
\partial_{\mu} U^{\dagger}) \label{nlsigma}
\end{eqnarray}
where
\begin{eqnarray} 
U=\exp(i \vec{\tau} \cdot \vec{\phi}/ f_{\phi})\label{u}
\end{eqnarray}
 
The classical soliton of Eq.(\ref{nlsigma}) has been found to be very 
important because of its usefulness in describing hadrons (Skyrmions)
\cite{skyrme} in 
low energy hadron physics as well as the approxmate behavior of the global
texture which is attracting interest as a potential source of density
perturbation for large scale structure of the universe \cite{texture}.  Since 
the simplest form of Eq.(\ref{nlsigma}) cannot support a stable soliton with 
finite
energy\cite{derrick}, higher 
derivative terms are needed in Eq.(\ref{lphi4}) or Eq.(\ref{nlsigma}) to
describe the system with finite energy. They are the Skyrme terms used in low 
energy hadron physics.

It is observed\cite{iwasaki} that there is a set of solutions which extremize 
the action with the simplest lagrangian density, Eq.(\ref{nlsigma}), using the
hedgehog ansatz:
\begin{eqnarray} 
U = \exp( i \vec{\tau} \cdot \hat{\vec{r}} F(r))\label{hedge}
\end{eqnarray}
The profile function F(r) satisfies the following equation;
\begin{eqnarray}
\frac{d^2 F}{dr^2} + \frac{2}{r} \frac{d F}{dr}=\frac{1}{r^2}\sin{2F}\label{f1}
\end{eqnarray}
In this equation, one can see that there is 
no scale parameter which can determine the size of the soliton. This leads 
to the conclusion that, due to
Derrick's  theorem, there is no stable solution with finite energy.
Fig.1 shows a numerical solution of Eq.(\ref{f1}) with the boundary condition
\begin{eqnarray}
F(0)=\pi\label{fzero}.
\end{eqnarray}
It was analytically proved in Ref.\cite{iwasaki} that there are no other 
solutions
except for those solutions that can be obtained through the following 
transformations;
\begin{eqnarray}
F(r) & \rightarrow & F(\lambda r) \ \ \ \ \ \ \  (\lambda =constant)
\nonumber \\
     & \rightarrow & F(r)+n\pi\ \            (n=integer)\nonumber \\
     & \rightarrow & -F(r) \label{trs}
\end{eqnarray}
The profile function $F(r)$ in Fig.1 approaches
$\pi /2$ as $r \rightarrow \infty$ as can be seen in the asymptotic form:
\begin{eqnarray} 
F(r) \rightarrow \frac{\pi}{2} + \sqrt{\frac{r_0}{r}} \cos \left( 
\frac{\sqrt{7}}{2} 
\ln(\frac{r_0}{r}) +\alpha \right) \label{half}
\end{eqnarray}
The winding number defined by
\begin{eqnarray}
N=\frac{F(0) - F(\infty)}{\pi} - \frac{\sin \, 2F(0) - \sin \, 2F(\infty)}
{2 \pi}
\end{eqnarray}
can be calculated to be $\frac{1}{2}$.
Although the energy of this solution is infinite, it cannot be excluded 
mathematically as one of  the solutions that extremize the action.
It might be useful in describing the physical situations where a very large 
number of particles are involved in heavy ion collisions or where the 
solution has to be truncated 
due to the finite size of the correlated volume which is restricted by the 
horizon in the early universe. In Ref.\cite{ttexture}, 
the time evolution of the textures whose initial configurations are given
by small perturbations from the solution in Fig.1 was investigated.
Within the one parameter set of initial configurations considered there,
it was found that textures with winding number larger(smaller) than 1/2
collapse(expand). The critical nature of the winding number 1/2 was also
discussed in Ref.\cite{lp}. The author presented the analytical predictions 
about the effect of the winding number on behaviors(time evolutions) of the
scalar field and showed configurations of the predictions by numerically 
solving the field equations with a couple of particular forms of initial
configurations.
Although the above findings do not provide the full general proof,
they lead one to suspect that the critical property that determines the 
collapse is whether the winding number is larger than 1/2. Ref.\cite{bcl}
 presents numerical investigations of the collapse and unwinding of
global textures without recourse to the nonlinear $\sigma$-model 
approximation. 
As was explicitly shown in Ref.\cite{ttexture}, the static solution shown
in Fig.1 is not a stable one. However, being a unique static solution that
extremizes the action, it may represent the most probable initial 
configuration of texture formation. A small perturbation will eventually cause
it to collapse or expand, but its collapse or expansion will be delayed longer
than that of any other initial configuration.
  
When we consider the situation with very large number of particles or with
very large amount of energy, the effects of gravity and/or quantum
corrections which can be put into the lagrangian in the form of higher 
derivative terms may become important. 
Hence it is interesting to investigate those effects on the 
winding number one-half solution.
Here we consider the effects of gravity    starting with the following action;
\begin{eqnarray}
I = \int d^4 x \sqrt{-g} \left(\frac{-1}{16\pi G}R-\frac{f_{\phi}^2}{4}
Tr(\partial^{\mu}U \partial_{\mu}U^{\dagger})\right) \label{action}
\end{eqnarray}
with $U$ given by Eq.(\ref{hedge}). For earlier works 
concerning the system of texture coupled with gravity, see for example Ref.'s
\cite{bv}, \cite{nz} and \cite{dhjs}.

For a static and spherically symmetric solution,
the metric can be written as   
\begin{eqnarray}
ds^2 = e^{2q(r)} dt^2 - e^{2p(r)} dr^2 - r^2 (d \theta ^2 + \sin^2\theta 
d\phi ^2).\label{line}
\end{eqnarray}
The matter  field equation obtained from $\delta I/\delta F=0$ is
\begin{eqnarray}
\frac{d}{dr}(r^2 \frac{dF}{dr})+ (\frac{dq}{dr}
- \frac{dp}{dr})r^2 \frac{dF}{dr}  - e^{2p} \sin(2F) =0\label{matter1}
\end{eqnarray}
The equations for $p(r)$ and $q(r)$ obtained from the Einstein's equations,
\(G^{\mu \nu}=-8\pi G \ T^{\mu \nu}\), are
\begin{eqnarray}
\frac{dp}{dr} = \frac{1}{2r}(1-e^{2p})+\frac{\kappa r}{4}\left( 
(\frac{dF}{dr})^2 + e^{2p} \frac{2 \sin^2 F}{r^2} \right)\label{goo}
\end{eqnarray}
\begin{eqnarray}
\frac{dq}{dr}=-\frac{1}{2r}(1-e^{2p})+\frac{\kappa r}{4}\left(
(\frac{dF}{dr})^2 - e^{2p} \frac{2 \sin^2 F}{r^2} \right)\label{goi}
\end{eqnarray}
where $\kappa = 8 \pi G f_{\phi}^2$.
From Eq's.(\ref{matter1}), (\ref{goo}) and ( \ref{goi} ), we can see there 
is no scale parameter, and hence no scale to determine the size of the 
solution. 
It is then an interesting question whether there exists  a static solution 
which extremizes the action, Eq.(\ref{action}), which corresponds to 
Eq.(\ref{half}) without gravity.
In our calculations, we first put Eq.(\ref{goi}) into Eq.(\ref{matter1})
to eliminate $q(r)$, couple the resulting equation with Eq.(\ref{goo})
and numerically integrate them to obtain $F(r)$ and $p(r)$ with the 
boundary condition
\begin{eqnarray}
F(0)=\pi, \ \ \ \ p(0)=0.\label{bound}
\end{eqnarray}  
The only other input in the calculation is the value of $dF/dr$ at $r=0.$
Changing the value induces only the change of the overall scale of the
solution. The result of the calculation is given in Fig.2.   
Here also a series of solutions can be generated from the solution in 
Fig.2 through the transformations similar to Eq.(\ref{trs}),
\begin{eqnarray}
F(r), p(r) & \rightarrow & F(\lambda r), p(\lambda r)\ \ \ \ (\lambda=constant)
                      \nonumber\\
           & \rightarrow & F(r) + n\pi, p(r) \ \ \ (n=integer)\nonumber\\
           & \rightarrow & -F(r), p(r) \label{trs2}       
\end{eqnarray}
Our numerical analysis indicates that
\begin{eqnarray}
\mid F(0) - F(\infty) \mid = \frac{\pi}{2}, \label{fdiff}
\end{eqnarray}
although we still lack the analytic proof.
The asymptotic forms at large $r$ can be calculated explicitly to be
\begin{eqnarray}
F &\rightarrow& \frac{\pi}{2} + \sqrt{\frac{r_0}{r}} \cos \left( 
\frac{\beta}{2} \ln(\frac{r_0}{r}) + \gamma\right), \label{ghalff}\\
p & \rightarrow& a + b \frac{r_0}{r} \sin \left( \beta 
\ln(\frac{r_0}{r}) + \gamma +\delta\right), \label{ghalfp}\\
q &\rightarrow& c - (1 + \frac{\beta^2}{4})\frac{\kappa r_{0}}{4r} -
 \frac{1}{1 + \beta^2}\frac{\kappa r_{0}}{4r} cos\left(\beta 
ln(\frac{r_{0}}{r})   +2\gamma \right),\nonumber \\
  &  & +\frac{\beta^3}{4(1+\beta^2)}\frac{\kappa r_{0}}{4r}sin\left(
     \beta ln(\frac{r_{0}}{r}) + 2\gamma \right) - b \frac{r_{0}}{r}sin\left(
       \beta ln(\frac{r_{0}}{r}) + \gamma + \delta \right)\label{ghalfq}  
\end{eqnarray}
where
\begin{eqnarray}
a &=& \frac{1}{2} \ln (\frac{1}{1-\kappa}),\label{ahalf}\\
\beta^2 &=&\frac{7+\kappa }{1-\kappa},\label{beta}\\
b^2 &=& \frac{1}{256} \frac{\kappa^2 }{1-\kappa},
\label{delta}\\
\tan \, \delta &=& - \beta,\label{phihalf}
\end{eqnarray}
and $r_0$, $\gamma$ and $c$ are integration constants. We can put $c$ to
be zero by adjusting the scale of the coordinate $t$ so that it coincides
with the scale of proper time in the asymptotic region of $ r \rightarrow 
\infty$.

Let us now consider the geometry of the hypersurface of $t$ = constant in the
asymptotic region of large $r$. The asymptotic form of the metric is given by
\begin{eqnarray}
ds^2=\frac{1}{1-\kappa}dr^2 + r^2(d\theta^2 + \sin^2\theta d\phi^2) 
\end{eqnarray}
or with the coordinate transformation, \(\tilde{r}=
\sqrt{\frac{1}{1-\kappa}}r\),
\begin{eqnarray}
ds^2=d\tilde{r}^2 + (1-\kappa)\tilde{r}^2 (d\theta^2 + \sin^2\theta d\phi^2)
\end{eqnarray}
This metric represents a flat space in the asymptotic region with the 
deficit solid angle
\begin{eqnarray}
\delta\Omega=4\pi\kappa
\end{eqnarray}
To see that the metric represents a flat space in the asymptotic region,
we calculate all components of the Riemann curvature tensor in the 
orthonormal basis which we take as
\begin{eqnarray}
e^1=\sqrt \frac{1}{1-\kappa} dr,\ \ \  e^2=rd\theta, \ \ \
 e^3=r \sin\theta d\phi
\end{eqnarray}
The only nonzero components are
\begin{eqnarray}
R^2_{\ 323}=-\frac{\kappa}{r^2}
\end{eqnarray}
and those related to it by algebraic symmetry or antisymmetry. 
We see that all components of the Riemann curvature tensor in the 
orthonormal basis converge to zero as $r\rightarrow \infty$.
It has been shown in Ref.\cite{missing} that the self-similar texture 
solution also
produces a flat spatial hypersurface with some missing solid angle in the
asymptotic region of large $r$.

We next consider the energetics of our solution. With the use of the hedgehog
ansatz (Eq.(\ref{hedge})) and the metric form of Eq.(\ref{line}), the 
$tt$-component of the energy-momentum tensor $T^{\mu \nu}$ is calculated to be
\begin{eqnarray}
T^{tt} = \frac{f_{\phi}^{2}}{4}\left\{e^{-2q-2p}\left(\frac{dF}{dr}\right)^2 +
e^{-2q}\frac{2 sin^2 F}{r^2} \right\}.\label{ttt}
\end{eqnarray}     
 To investigate the behavior($r$ - dependence) of the energy density in the 
asymptotic flat region of $r \rightarrow \infty$, we put the asymptotic
solutions, Eq.'s (\ref{ghalff}), (\ref{ghalfp}) and (\ref{ghalfq}), into
Eq. (\ref{ttt}) and obtain the result
\begin{eqnarray}
T^{tt} \sim \frac{1}{r^2}
\end{eqnarray}
as $r \rightarrow \infty$. This shows that the total energy contained in a
volume diverges linearly in $r$ as the volume is increased.
 
In conclusion, we see that the basic structure of the classical static 
solution is not changed by gravity: there is still no scale parameter that
determines the size of the field configuration and the solution represents
the knot of windung number $1/2$. The geometry of the spatial hypersurface 
in the asymptotic region of large $r$ is that of a flat space with some 
missing solid angle. 

\acknowledgements
This work was supported
in part by the KOSEF under Grant No.91-08-00-04 and by the Korean 
Ministry of Education (BSRI-93-231)

\figure{The profile function $F(r)$ without gravity.}
\figure{The profile function $F(r)$ and the function $p(r)$ with gravity 
($\kappa = 0.865$).}

\begin{references}
\bibitem{goldstone}
J. Goldstone, Nuovo Cim. {\bf 19}, 154 (1991); Y. Nambu and G. Jona-Lasinio,
Phys. Rev. {\bf 122}, 154 (1961)
\bibitem{skyrme}
T. H. R. Skyrme, Proc. Roy. Soc. London {\bf A260}, 127 (1961); Nucl. Phys.
{\bf 31}, 556 (1962)
\bibitem{texture}
R. Davis, Phys. Rev. {\bf D35}, 3705 (1987); {\bf D36}, 997 (1987); N. Turok,
Phys. Rev. Lett. {\bf 63}, 2625 (1989)
\bibitem{derrick}
G. H. Derrick, J. Math. Phys. {\bf 5}, 1252 (1964)
\bibitem{iwasaki}
M. Iwasaki and H. Ohyama, Phys. Rev. {\bf D40}, 3125 (1989)
\bibitem{ttexture}
S. Aminneborg, Stockholm Univ. preprint USITP-91-21 (1992)
\bibitem{lp}
L. Perivolaropoulos, Phys. Rev. {\bf D46}, 1858(1992)
\bibitem{bcl}
J. Borrill, E. Copeland and A. Liddle, Phys. Rev. {\bf D46}, 524(1992): Phys.
 Rev. {\bf D47}, 4292(1993)
\bibitem{bv}
M. Barriola and T. Vachaspati, Phys. Rev. {\bf D43}, 1056(1991): Phys. Rev.
 {\bf D43}, 2726(1991)   
\bibitem{nz}
D. N\"{o}tzold, Phys. Rev. {\bf D43}, R961(1991)
\bibitem{dhjs}
R. Durrer, M. Heusler, P. Jelzer and N. Struamann, Nucl. Phys. {\bf B368}, 
527(1992)
\bibitem{missing}
N. Turok and D. Spergel, Phys. Rev. Lett. {\bf 64}, 2736 (1990) 
\end{references}
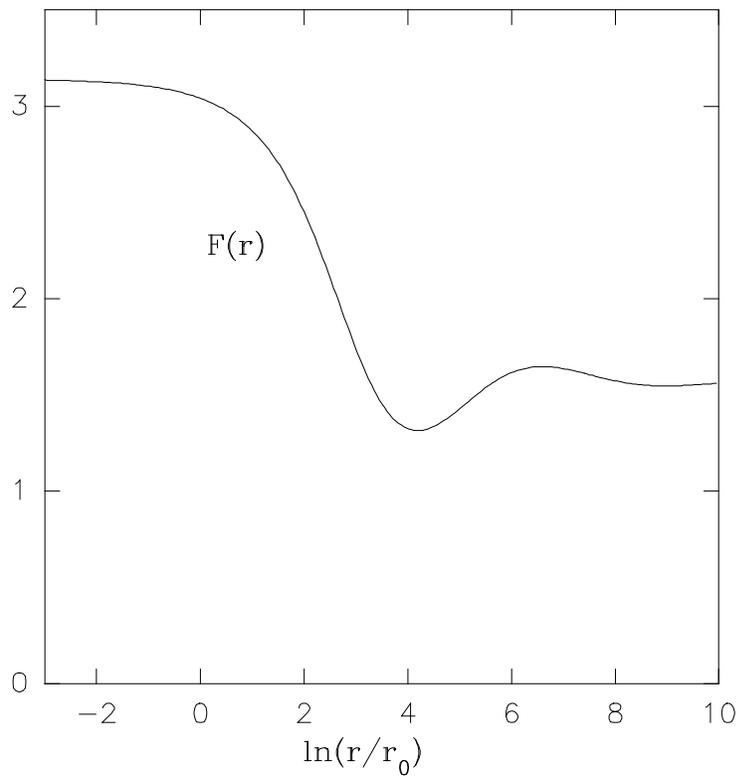
\end{document}